\documentclass[12pt,a4paper]{article}
\makeatletter
\newcommand*{\indep}{%
  \mathbin{%
    \mathpalette{\@indep}{}%
  }%
}
\newcommand*{\nindep}{%
  \mathbin{
    \mathpalette{\@indep}{/}%
  }%
}
\newcommand*{\@indep}[2]{%
  \sbox0{$#1\perp\m@th$}
  \sbox2{$#1=$}
  \sbox4{$#1\vcenter{}$}
  \rlap{\copy0}
  \dimen@=\dimexpr\ht2-\ht4-.2pt\relax
  \kern\dimen@
  \ifx\\#2\\%
  \else
    \hbox to \wd2{\hss$#1#2\m@th$\hss}%
    \kern-\wd2 %
  \fi
  \kern\dimen@
  \copy0 
}
\makeatother
\usepackage{rotating}
\usepackage{float}
\usepackage{siunitx}
\usepackage{mathrsfs}  
\usepackage[latin1]{inputenc}
\usepackage{amsmath}
\usepackage{amssymb}
\usepackage{amsfonts}
\usepackage{graphicx}
\usepackage{color}
\usepackage[numbers]{natbib}
\usepackage{array} 
\usepackage{mathrsfs}  
\usepackage{url}  
 \usepackage{mathabx}
\usepackage{pgf,tikz}
\usetikzlibrary{shapes.geometric, positioning}
\usepackage{booktabs} 
\usepackage{colortbl} 
\usepackage{xcolor} 
\usepackage{xfrac}
\usepackage{makecell}
\usetikzlibrary{cd}
\newcommand{\blind}{1}
\begin{document}

\def\spacingset#1{\renewcommand{\baselinestretch}%
{#1}\small\normalsize} \spacingset{1}
\date{ }



\if1\blind

{
\title{\bf Comparing the  effectiveness of $k$-different treatments through the area under the ROC curve} 
\author{Pablo Mart\'inez-Camblor$^{1,2}$\thanks{{\it Correspondence to:} Pablo Mart\'inez-Camblor. 7 Lebanon Street, Suite 309, Hinman Box 7261, Lebanon, NH 03751, USA. E-mail: {\color{blue} Pablo.Martinez-Camblor@hitchcock.org}}, Sonia P\'erez-Fern\'andez$^3$,\\ Lucas L. Dwiel$^4$ and Wilder T. Doucette$^{4}$\\
\hspace{.2cm}\\
\small{$^1$Department of Anesthesiology, Geisel School of Medicine at Dartmouth, NH, USA}\\ 
\small{$^2$Faculty of Health Sciences, Universidad Autonoma de Chile, Chile}\\
\small{$^3$Department of Statistics and O.R. and M.D.,  Oviedo University, Asturies, Spain}\\
\small{$^4$Department of Psychiatry, Geisel School of Medicine at Dartmouth, NH, USA}}
\maketitle
} \fi

\begin{abstract}
The area under the receiver-operating characteristic curve (AUC) has become a popular index not only for measuring the overall prediction capacity of a marker but also the strength of the association between continuous and binary variables. In the current study, the AUC was used for comparing the association size of four different interventions involving impulsive decision making, studied through an animal model, in which each animal provides several negative (pre-treatment) and positive (post-treatment) measures. The problem of the full comparison of the average AUCs arises therefore in a natural way. We construct an analysis of variance (ANOVA) type test for testing the equality of the impact of these treatments measured through the respective AUCs, and considering the random-effect represented by the animal. The use (and development) of a post-hoc Tukey's HSD type test is also considered. We explore the finite-sample behavior of our proposal via Monte Carlo simulations, and analyze the data generated from the original problem. An R package implementing the procedures is provided as supplementary material.
\end{abstract}
\vfill

\spacingset{1.45} 
\section{Introduction}
Impulsivity is a multi-dimensional psychological domain describing the predisposition toward rapid, unplanned reactions to internal or external stimuli with diminished regard to the negative consequences of these reactions to the individual or to others. Impulsivity is further subdivided into two distinct types: 1) impulsive actions related to refraining from initiating an action or stopping an action that has been initiated; and 2) impulsive decisions related to a lack of planning or lack of regard for future consequences. Maladaptive levels of impulsivity have been associated with several psychiatric disorders (e.g., bipolar disorder, substance use disorders, and  personality disorders) \cite{moeller01}. In humans, different dimensions of impulsivity are quantified through well-established self-report measures as well as performance on specific tasks. In rodents, related tasks are used to assess different domains of impulsivity. For example, the five-choice serial reaction-time task (5-CSRTT) is commonly used to assess impulsive actions and the delay discounting task (DDT) is used to assess impulsive decisions \cite{mitchell14}. Across species, various techniques are used to find neural biomarkers of impulsivity through quantifications of brain activity like the blood oxygenation level dependent (BOLD) signal from functional magnetic resonance imaging (fMRI) or from electrophysiological methods like electroencephalography (EEG)  and local field potentials (LFPs).

Using quantifications of brain activity to compare the magnitude and nature of brain activity changes induced by various interventions in a feature-agnostic fashion quickly falls outside the scope of traditional statistical methods such as t-tests, as the number of features describing brain activity increases. We quantify brain activity using power and coherence extracted from local field potentials (LFPs) across a range of frequencies and brain regions (bilateral orbitofrontal cortex, infralimbic cortex, and nucleus accumbens core and shell), providing 216 features (see \citet{dwiel19} for signal processing details). Briefly, coherence quantifies the degree to which two signals correlate with one another at a given frequency, in this case providing a metric of connectivity between pairs of brain regions while power reflects the amount of a given frequency present in a signal recorded from a single brain region \cite{budsaki12, harris17}. By collecting many measures both before and after an intervention, we can then use the penalized logistic regression model using the least absolute shrinkage and selection operator (LASSO) \cite{tibshirani96} and then using the area under the receiver operator characteristic curve, AUC \cite{hanley82}, as a metric to represent the difference in brain states induced by an intervention. Then, to compare the change in brain states across multiple interventions (brain stimulation of infralimbic cortex, brain stimulation of nucleus accumbens core, an injection of methylphenidate, or an injection of saline) motivated this work. Particularly, we focus here in the methodological aspects of the last part of the analysis. It involves evaluating and comparing the impact of the four different treatments using measures obtained in clusters defined by the animal. The aim of this paper is to construct an analysis of variance (ANOVA) type test for testing the equality of the impact of these treatments measured through the respective AUCs, and considering the random-effect represented by the animal.

Rest of the paper is organized as follows. In Section 2, we revise asymptotic properties of the empirical estimator for the AUC. These properties are used in Section 3 in order to develop our proposal. In Section 4, the post hoc test based on the Tukey's HSD procedure is developed. Finite-sample behavior of the proposed methodology, including the post hoc test, is studied in Section 5 via Monte Carlo simulations (as supplementary material, we provide more simulation results). In Section 6, we fully analyze the dataset which motivated this research. Highlighting that, since the original research is still in process, data presented have been manipulated, and no (real) neuroscience specific conclusions should be derived from our analyses. In Section 7, we make a brief presentation of the R package used for implementing our proposal, and available at \url{https://github.com/perezsonia/aovAUC}. Finally, in Section 8, we present our main conclusions. Some technical details were relegated to Appendix.

\section{General framework and notation}
Let $\xi$ and $\chi$ be two continuous and independent random variables modeling the behavior of the consider biomarker in the positive and in the negative populations, respectively. Let $F_\xi(\cdot)$ and $G_\chi(\cdot)$ be their respective cumulative distribution functions (CDFs). Then, assuming (without loss of generality) that $\mathbb E[\xi]\geq\mathbb E[\chi]$, the area under the ROC curve, AUC, is determined by
\begin{equation}
{\cal A}=1 - \int F_\xi(x)dG_\chi(x)={\cal P}\{\chi<\xi\}.
\end{equation}
Let $\{y_1,\cdots ,y_n\}$ and $\{x_1,\cdots , x_m\}$ be two independent random samples drawn from $\xi$ and $\chi$, respectively. The empirical AUC estimator \cite{hanley82} is the resulting of replacing the unknown CDFs by their maximum-likelihood estimators (ECDFs). That is,
\begin{equation}
\hat {\cal A}_n = \frac{1}{n\cdot m}\sum_{i=1}^n\sum_{j=1}^m \left\{ I(x_j < y_i) + \frac{1}{2}I(x_j=y_i)\right\},
\end{equation}
where $I(A)$ is the standard indicator function (takes the value $1$ if $A$ is true and $0$ otherwise). Both theoretical properties \cite{pepe03, zhou02}, and practical use \cite{janssens20,dehond22} of the AUC have been deeply considered in the specialized literature. 
Besides, the AUC has been proposed as alternative index for measuring the association between continuous (or ordinal) covariates and binary outcomes \cite{demidenko16, camblor20}. 

If $n/m=\lambda_n\longrightarrow_n\,\lambda >0$, then the projection (see, for instance, \citet{van00}, Chapter 12), 
\begin{equation*}
\mathbb P_{\hat {\cal A}_n}= \left\{\frac{1}{m}\sum_{i=1}^m F_\xi(x_i) - \mathbb E[F_\xi(\chi)]\right\} - \left\{\frac{1}{n}\sum_{i=1}^n G_\chi(y_i) - \mathbb E[G_\chi(\xi)] \right\},
\end{equation*}
satisfies the weak convergence
\begin{equation*}
\sqrt{n}\cdot \mathbb P_{\hat {\cal A}_n}\stackrel{\cal L}{\longrightarrow}_n\, \mathscr N(0,\sigma), 
\end{equation*}
where $\sigma^2= \lambda\cdot \| F_\xi\cdot G^{-1}_\chi \|^* + \|G_\chi\cdot F^{-1}_\xi\|^*$, with $\|h\|^*=\int_0^1 h^2(t)dt - \left(\int_0^1 h(t)dt\right)^2$ \cite{hsieh96}. In practice, $\sigma^2$ can be approximated replacing the unknown involved CDFs by their respective ECDFs, and $\lambda$ for $\lambda_n$. Besides,
\begin{equation*}
\sqrt{n}\cdot\left\{ \mathbb P_{\hat {\cal A}_n} - [\hat {\cal A}_n - \cal A]\right\} \stackrel{\cal P}{\longrightarrow}_n\, 0.
\end{equation*}
Therefore, the Slutsky's Lemma guarantees the weak convergence
\begin{equation}
\sqrt{n}\cdot [\hat {\cal A}_n - {\cal A}] \stackrel{{\cal L}}{\longrightarrow}_n\, {\mathscr N}(0,\sigma).\label{weak}
\end{equation}
The comparison of two AUCs derived from independent samples is immediate from the above approximation and the normal distribution properties. Comparisons involving related samples have also been already considered in the specialized literature \cite{delong88, bandos05}.

\section{ANOVA-type test for AUCs comparison}

In our problem (see the Introduction section), we have animals independently assigned to $k$ ($=4$, in our particular case) treatments. Each subject has a number of pre-treatment (control) and a number of post-treatment (case) measures. Since the AUC is used for measuring the impact of the treatment, and we are interested in comparing the difference between them, we have to test the null
\begin{equation}
H_0: {\cal A}_1=\cdots ={\cal A}_k\quad (={\cal A}), \label{null}
\end{equation}
where ${\cal A}_i$ ($1\leq i \leq k$) is the AUC associated with the $i$-th treatment. Let $\{y_{i,j,1}, \cdots , y_{i,j,m_{P_{i,j}}}\}$ and $\{x_{i,j,1}, \cdots , x_{i,j,m_{N_{i,j}}}\}$ be the $m_{P_{i,j}}$ post-treatment and the $m_{N_{i,j}}$  pre-treatment measures of the $j$-th animal ($1\leq j\leq n_i$) receiving the $i$-th treatment ($1\leq i \leq k$), and let $\hat {\cal A}_{i,j}$ be the empirical AUC estimator associated with this animal. Assuming $m_{P_{i,j}}/m_{N_{i,j}}\longrightarrow_{m_{P_{i,j}}}\, \lambda_{i,j}>0$, we have that the distribution of $[\hat {\cal A}_{i,j} - {\cal A}_{i,j}]$ can be approximated through a $\mathscr N(0,  \hat\sigma_{i,j}[n_i])$, with 
$$\hat\sigma^2_{i,j}[n_i]= m_{N_{i,j}}^{-1}\cdot \| \hat F_{m_{P_{i,j}}}\cdot \hat G^{+}_{m_{N_{i,j}}}\|^* + m_{P_{i,j}}^{-1}\cdot \| \hat G_{m_{N_{i,j}}}\cdot \hat F^{+}_{m_{P_{i,j}}}\|^*,$$
\noindent
where $\hat F_{m_{P_{i,j}}}(\cdot)$ and $\hat G_{m_{N_{i,j}}}(\cdot)$ are the ECDFs referred to the samples $\{y_{i,j,1}, \cdots , y_{i,j,m_{P_{i,j}}}\}$ and $\{x_{i,j,1}, \cdots , x_{i,j,m_{N_{i,j}}}\}$, respectively, 
$\hat F^{+}_{m_{P_{i,j}}}(\cdot)=\{t: \inf \hat F_{m_{P_{i,j}}}(t)\geq \cdot\}$, and $\hat G^{+}_{m_{N_{i,j}}}(\cdot)=\{t: \inf \hat G_{m_{N_{i,j}}}(t)\geq \cdot\}$. Assuming that, for $j\in\{1,\cdots , n_i\}$, ${\cal A}_{i,j} = {\cal A}_i + a_{i,j}$, where $a_{i,j}$ is the individual effect of the $j$-th subject when they receives the $i$-th treatment ($1\leq i\leq k$), and that within each treatment, this individual effect has mean zero, $\mathbb E[a_{i,\bullet}]=0$, and (between-subjects) variance $\tau^2_{a_{i,\bullet}}$ ($=\mathbb V[a_{i,\bullet}]$), the distribution of the $n_i$-dimensional random vector
\begin{align*}
[\boldsymbol {\hat {\cal A}_{i,\bullet} - {\cal A}_i}]=&\{[\hat {\cal A}_{i,1} -{\cal A}_i], \cdots, [\hat {\cal A}_{i,n_{i}} -{\cal A}_i]\}\\
                                                                            =&\{[\hat {\cal A}_{i,1} -{\cal A}_{i,1}] + a_{i,1}, \cdots, [\hat {\cal A}_{i,n_{i}} -{\cal A}_{i,n_i}] + a_{i,n_i}\}, 
\end{align*}
can be approximated through a $\mathscr N_{n_i}(\boldsymbol 0, \boldsymbol {\Sigma(n_i)})$ distribution, where the variance-covariance matrix is determined by
$$\boldsymbol {\Sigma(n_i)}= \{\hat\sigma^2_{i,1}[n_i] + \hat\tau^2_{a_{i,\bullet}},\cdots , \hat\sigma^2_{i,n_i}[n_i]+ \hat\tau^2_{a_{i,\bullet}}\}\cdot \boldsymbol {I_{n_i}},$$
where $\hat\tau^2_{a_{i,\bullet}}$ is an estimation of the between-subjects variance (more information about this estimation is provided as appendix) under the $i$-th treatment, and $\boldsymbol {I_{n_i}}$ stands for the $n_i\times n_i$ identity matrix. Hence, given the $n_i\times n_i$ symmetric matrix
\begin{equation*}
\boldsymbol {U_{n_i}}= \boldsymbol {I_{n_i}} - \frac{1}{n_i}  \begin{bmatrix} 1 & 1 & \dots & 1 \\     \vdots & \vdots & \ddots & 1 \\  1 & 1 & \dots    & 1 
    \end{bmatrix},
\end{equation*}
the distribution of $[\boldsymbol {\hat {\cal A}_{i,\bullet} - {\cal A}_i}]\cdot \boldsymbol {U_{n_i}}$ can be approximated by a $\mathscr N_{n_i-1}(\boldsymbol 0, \boldsymbol {S(n_i)})$ distribution, where $\boldsymbol {S(n_i)}=\boldsymbol {U_{n_i}}\cdot \boldsymbol {\Sigma(n_i)}\cdot \boldsymbol {U_{n_i}}$. Since $\boldsymbol {\hat {\cal A}_i}\cdot \boldsymbol {U_{n_i}}=\boldsymbol 0$, the distribution of $\boldsymbol {\hat {\cal A}_{i,\bullet}}\cdot  \boldsymbol {S^+(n_i)}\cdot \boldsymbol {\hat {\cal A}^t_{i,\bullet}}$, where  $\boldsymbol {S^+(n_i)}$ is the Moore-Penrose inverse  matrix of  $\boldsymbol {S(n_i)}$, can be approximate by a $\chi^2_{n_i -1}$ distribution. Therefore, the distribution of the sum of the intra-groups variability
\begin{equation}
\text{SSE}= \sum_{i=1}^k \boldsymbol {\hat {\cal A}_{i,\bullet}}\cdot  \boldsymbol {S^+(n_i)}\cdot \boldsymbol {\hat {\cal A}^t_{i,\bullet}}
\end{equation}
can be approximated through a $\chi^2_{n - k}$, where $n=\sum_{i=1}^k n_i$ is the total number of subjects included in the study. \par
\vspace{0.4cm}
On the other hand, for $i\in \{1,\cdots , k\}$,
\begin{equation*}
\hat {\cal A}_{i,\bullet }=\frac{1}{n_i}\sum_{i=1}^{n_i} \hat {\cal A}_{i,j}
\end{equation*}
is a consistent estimator for ${\cal A}_i$, which satisfies the convergence
\begin{equation*}
\sqrt{n_i}\cdot [\hat {\cal A}_{i,j} - {\cal A}_i] \stackrel{{\cal L}}{\longrightarrow}_n\, {\mathscr N}(0, \sigma_{i,\bullet}),
\end{equation*}
with $\sigma^2_{i,\bullet}=n_i^{-1}\cdot \sum_{i=1}^{n_i}[\sigma^2_{i,j} + \tau^2_{a_{i,\bullet}}]$, and where $\sigma^2_{i,j}$ ($1\leq j\leq n_i$) is the asymptotic variance derived from the convergence provided in (\ref{weak}), and associated with the $j$-th subject in the $i$-th treatment, and $\tau^2_{a_{i,\bullet}}$ is the between-subjects variance. Following the previous argument, when the null hypothesis in (\ref{null}) is true, the distribution of the random vector
\begin{equation*}
[\boldsymbol {\hat {\cal A}_{\bullet,\bullet} - {\cal A}}]=\{[\hat {\cal A}_{1,\bullet} -{\cal A}], \cdots, [\hat {\cal A}_{k,\bullet} -{\cal A}]\},
\end{equation*}
can be approximated through a ${\mathscr N}_k(\boldsymbol 0, \boldsymbol {\Sigma(k)})$ distribution, with 
$$\boldsymbol {\Sigma(k)}= \{\hat\sigma^2_{1,\bullet}[k],\cdots , \hat\sigma^2_{k,\bullet}[k]\}\cdot \boldsymbol {I_{k}},$$
where, for $1\leq i\leq k$, $\hat\sigma^2_{i,\bullet}[k]$ is the natural estimator for $\sigma^2_{i,\bullet}$ based on the above mentioned approximations for $\sigma^2_{i,j}$ ($1\leq j\leq n_i$) and $\tau^2_{a_{i,\bullet}}$, ($\hat\sigma^2_{i,j}[n_i]$ and $\hat\tau^2_{a_{i,\bullet}}$, respectively).

Therefore, under the null hypothesis (\ref{null}), the distribution of $[\boldsymbol {\hat {\cal A}_{\bullet,\bullet} - {\cal A}}]\cdot \boldsymbol {U_{k}}$ can be approximated by $\mathscr N_{k}(\boldsymbol 0, \boldsymbol {S(k)})$, where $\boldsymbol {S(k)}=\boldsymbol {U_{k}}\cdot \boldsymbol {\Sigma(k)}\cdot \boldsymbol {U_{k}}$. And arguing as previously, the distribution of the sum of the inter-groups variability
\begin{equation}
\text{SSF}= \boldsymbol {\hat {\cal A}_{\bullet,\bullet}}\cdot  \boldsymbol {S^+(k)}\cdot \boldsymbol {\hat {\cal A}^t_{\bullet,\bullet}}
\end{equation}
can be approximated through a $\chi^2_{k-1}$ distribution. We propose to use the adjusted quotient among (5) and (6) for testing the target null, since we know that
\begin{equation}
F_{k-1, n-k}= \frac{(n-k)\cdot \text{SSF}}{(k-1)\cdot\text{SSE}}
\end{equation}
can be approximate through a F-Snedecor distribution with $k-1$ and $n-k$ degrees of freedom, and that larger values of this quotient would indicate less likelihood of the null hypothesis being true.

\section{Tukey's HSD-type test}
The procedure described above allows to compute a p-value useful for making decisions related with the null (\ref{null}). However, if we decide to reject this null, that is, to assume that not all the AUCs in the considered groups are equal, we do not know where these differences are located. Standard  analyses of variance are frequently complemented with post hoc comparisons between all the potential pairs. The main handicap is that with the number of involved groups, the number of pairs to compare drastically increments, and with this, the multiple comparison problem arises. Post hoc tests \cite{holm85} provide full pair comparisons while pursuit to respect the fixed nominal level. The statistical literature is rich in post hoc tests proposals. Some of them are useful when the sample sizes in the different groups are similar, others do not assume homoscedasticity, Bonferroni method directly adjusts the nominal level by the number of pairs, among other particularities. We are adapting the well-known Tukey's Honest Significant Difference (HSD) test \cite{tukey49} to the AUC context.

The classical Tukey's HSD test computes the $k\cdot(k-1)/2$ so-called studentized difference among the $k$ means referred to the $k$ different groups, and uses the parametric assumption --traditional analyses of variance assume that we have underlying normal distributed random variables-- for computing the distribution of the maximum studentized difference. This distribution, under the null (all means are equal), so-called studentized range distribution, is tabulated and therefore each difference can be compared with the expected value when the null is true. For each individual pair we can compute a p-value (or a confidence interval for the studentized difference), which globally keep the fixed nominal level.

\begin{figure}
\begin{center}
{\includegraphics[width=12cm]{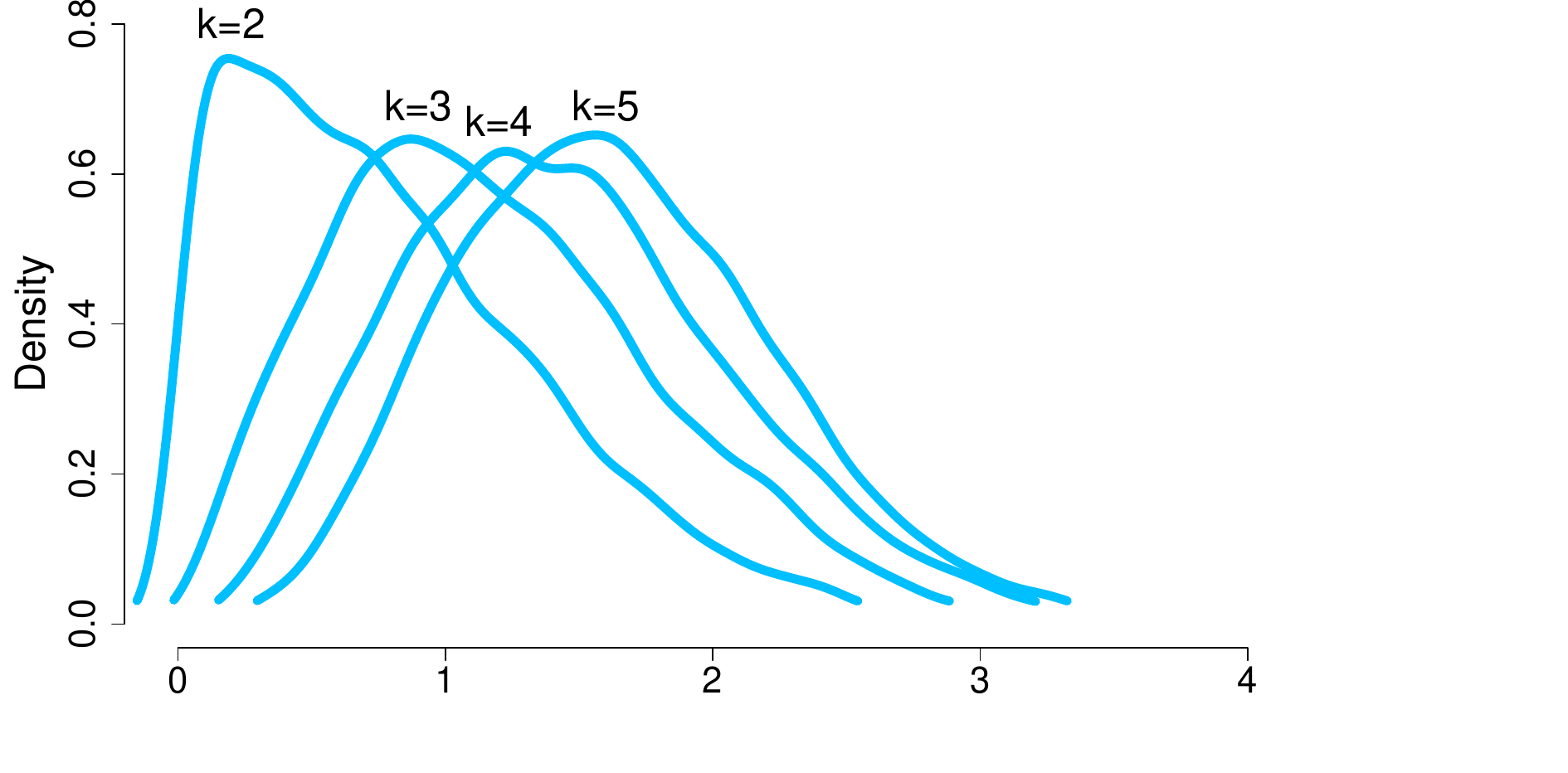}}\\
\end{center}
\caption{{\bf Post hoc distributions.} Kernel density estimations for the $mD[k]$ for $R=5000$, and for $k=2,\, 3,\, 4$, and 5. }
\label{posthoc}
\end{figure}

In the current context, we know that, under the null, the difference between the pairs,

$$\hat \Delta_{i,j}[k]=\frac{\hat {\cal A}_{i,\bullet} - \hat {\cal A}_{j,\bullet} }{\sqrt{\hat\sigma^2_{i.\bullet}[k] + \hat\sigma^2_{j.\bullet}[k] }}\quad (1\leq i\neq j\leq k)$$
can be approximated through a $\mathscr N(0,1)$ distribution. Therefore, the behavior of the random variable $m\hat\Delta[k]=\max_{1\leq i\neq j\leq k} |\hat \Delta_{i,j}[k]|$ can be approximated through the following algorithm.
\begin{enumerate}
\item[1.] Generate $R$ ($R$ a large enough number) values from $k$ independent normal distributed variables with mean zero and standard deviation 1, $(V_{i,1}, \cdots , V_{i,R})$, ($1\leq i \leq k$).
\item[2.] For $1\leq r \leq R$, compute $mD^r[k] = \max_{1 \leq i\neq j\leq k} |V_{i,r} - V_{j,r}|/\sqrt{2}$.
\item[3.] The distribution of $m\hat\Delta[k]$ is approximated through the values $(mD^1[k], \cdots , mD^R[k])$.
\end{enumerate}
The procedure compares each pair with this approximated distribution, which only depends on the number of groups. Both p-values and confidence intervals for the differences could be derived from the above algorithm. Figure \ref{posthoc} shows kernel density estimations based on $10^7$ points for $k=2,\,3,\, 4,\, 5$. In these cases, the thresholds determining the significance (at level $\alpha=0.05$) would be 1.961, 2.344, 2.569, and 2.727, respectively.

It is worth mentioning that, like the standard Tukey's HSD test, since we compare each pair value with the potential values of the maximum difference under the null, we anticipate a conservative behavior of the test.

\section{Monte Carlo simulations}
The finite sample size behavior of the ANOVA-type analysis proposed in (7) is studied through Monte Carlo simulations. Because the large number of involved parameters, and therefore the large number of different scenarios to consider, most part of our simulations are provided in the online supplementary material. We consider here an scenario with $4$ ($=k$) treatments, the number of animals per treatment was $r_{n_{R,i}}+2$ ($1\leq i \leq k$), where $r_{n_{R,i}}$ was generated from a Poisson distribution with parameter $n_R-2$, Pois($n_R-2$), and therefore, the expected and minimum numbers of rats per treatment were $n_R$  and 2, respectively. For each animal, we run the number of pre-treatment and post-treatment measures from two independent Poisson distributions with parameters $m_N-2$ and $m_P-2$, respectively, and then we add again 2  to the resulting numbers in order to have expected numbers of negative and positive measures per animal of $m_N$ and $m_P$, respectively. The negative values were always run from a standard normal distribution, $\mathscr N(0,1)$, while the positive values were run from $\mathscr N(\mu_i + \epsilon_{i,j}, 1)$, where $\mu_i$ ($1\leq i\leq k$) was chosen for having a treatment effect of ${\cal A}_i$ (=0.65, 0.75, and 0.85), and $\epsilon_{i,j}$ ($1\leq j\leq r_{n_R,i}+2$) stands for the random-effect of the subject, and it was generated from $\mathscr N(0, \sigma_\epsilon)$ (we considered $\sigma_\epsilon=0.3,\, 0.6$). All reported quantities are based on 5000 Monte Carlo iterations.

Table \ref{simul01} shows the rejection percentages when the nominal level is fixed to the usual 5\% (AV), and the mean$\pm$standard deviation number of pairs adequately rejected/no rejected (PH) when the four treatments have the same effect (${\cal A}$). Observed results suggest that the proposed approximation works adequately. Although, in general, it was  a little bit conservative (29 out 48 percentages were below 5\%, only 2 below 4\%). It seems that it could become anti-conservative if the sample sizes ($n_R$) is too small for handling the between-subjects variability (notice cases of $\sigma_\epsilon=0.6$, ${\cal A}=0.85$, $n_R=25$, and $m_N=m_P$, with rejection percentages larger than 6\%). The average number of no-rejected pairs (in the current case, the 6 possible pairs are equals) is very close to 5.95 indicating the accuracy of the used approximation (notice that if the 5\% of the samples wrongly reject 1 pair, the average would be 5.95). The percentage of samples which (erroneously) reject the equality among one or more pairs oscillated between 4.74 and 6.91 (average of 5.52\%), which also supports the correct behavior of the proposed post hoc test. 

\begin{footnotesize}
\begin{table}
\caption{{\footnotesize {\bf Null Hypothesis.} Percentage of rejections (AV) for $\alpha=0.05$, and mean$\pm$standard deviation of the number of pairs comparison success (PH) when the null is true computed from 5000 Monte Carlo iterations. Two expected number of subjects per group $n_R$, four combinations of the average numbers of expected pre-treatment ($m_N$) and post-treatment ($m_P$) measures, and three different AUCs (${\cal A}=0.65,\, 0.75,\, 0.85$) were considered.\label{simul01}}}
\begin{center}
\begin{footnotesize}
\hspace{-2cm}
\begin{tabular}{ccccccccccccccccccccccccc}
\rowcolor{black!15} 
& & && \multicolumn{5}{c}{$\boldsymbol {n_R=25}$} && \multicolumn{5}{c}{$\boldsymbol {n_R=50}$} \\ 
\rowcolor{black!15} 
& & && \multicolumn{2}{c}{$\boldsymbol \sigma_\epsilon=0.3$} &&  \multicolumn{2}{c}{$\boldsymbol  \sigma_\epsilon=0.6$} &&  \multicolumn{2}{c}{$\boldsymbol \sigma_\epsilon=0.3$} &&  \multicolumn{2}{c}{$\boldsymbol \sigma_\epsilon=0.6$}\\
\rowcolor{black!15} 
$\boldsymbol {m_N}$ & $\boldsymbol {m_P}$ & $\boldsymbol {\cal A}$  &&  {\bf AV} & {\bf PH} &&  {\bf AV} & {\bf PH} &&  {\bf AV} & {\bf PH} &&  {\bf AV} & {\bf PH}\\
25 & 25 & 0.65 &&  4.80 & 5.92$\pm$0.32 && 4.81 & 5.93$\pm$0.32 && 4.10 & 5.94$\pm$0.28 && 4.86 & 5.93$\pm$0.31\\ 
\rowcolor{black!5} 
   & 50 & 0.65 &&   4.82 & 5.92$\pm$0.33 && 5.99 & 5.91$\pm$0.35 && 4.38 & 5.94$\pm$0.30 && 5.22 & 5.93$\pm$0.30\\  
 50 & 50 & 0.65 && 5.72 & 5.92$\pm$0.35 && 5.32 & 5.93$\pm$0.31 && 4.42 & 5.94$\pm$0.30 && 4.94 & 5.94$\pm$0.30\\ 
\rowcolor{black!5}
    & 100 & 0.65 && 5.28 & 5.93$\pm$0.32 && 5.32 & 5.93$\pm$0.31 && 4.70 & 5.94$\pm$0.29 && 5.07 & 5.93$\pm$0.32\\  
25 & 25 & 0.75 && 4.02 & 5.93$\pm$0.31 && 5.94 & 5.91$\pm$0.34 && 4.14 & 5.93$\pm$0.30 && 4.36 & 5.94$\pm$0.28\\   
\rowcolor{black!5} 
    & 50 & 0.75 && 4.88 & 5.93$\pm$0.31 && 5.64 & 5.92$\pm$0.33 && 4.16 & 5.93$\pm$0.32 && 4.60 & 5.94$\pm$0.30\\   
50 & 50 & 0.75 && 4.92 & 5.93$\pm$0.30 && 5.40 & 5.93$\pm$0.32 && 3.88 & 5.94$\pm$0.30 && 4.98 & 5.93$\pm$0.30\\  
\rowcolor{black!5} 
   & 100 & 0.75 && 5.06 & 5.93$\pm$0.32 && 5.58 & 5.92$\pm$0.33 && 4.98 & 5.93$\pm$0.32 && 4.78 & 5.93$\pm$0.31\\  
25 & 25 & 0.85 && 3.92 & 5.94$\pm$0.30 && 6.12 & 5.93$\pm$0.33 && 4.10 & 5.94$\pm$0.30 && 5.66 & 5.94$\pm$0.29\\  
\rowcolor{black!5} 
    & 50 & 0.85 &&  4.90 & 5.92$\pm$0.32 && 4.42 & 5.93$\pm$0.31 && 4.32 & 5.94$\pm$0.30 && 4.78 & 5.94$\pm$0.29\\ 
50 & 50 & 0.85 && 5.18 & 5.93$\pm$0.33 && 6.24 & 5.92$\pm$0.32 && 4.52 & 5.94$\pm$0.28 && 5.50 & 5.93$\pm$0.30\\ 
\rowcolor{black!5} 
   & 100 & 0.85 && 4.68 & 5.93$\pm$0.30 && 5.72 & 5.93$\pm$0.31 && 4.80 & 5.93$\pm$0.31 && 5.04 & 5.94$\pm$0.29     
\end{tabular}
\hspace{-2cm}
\end{footnotesize}
\end{center}
\end{table}
\end{footnotesize}

\begin{figure}
\begin{center}
\begin{tabular}{cc}
{\bf Model 1.}\\
{\includegraphics[width=15cm]{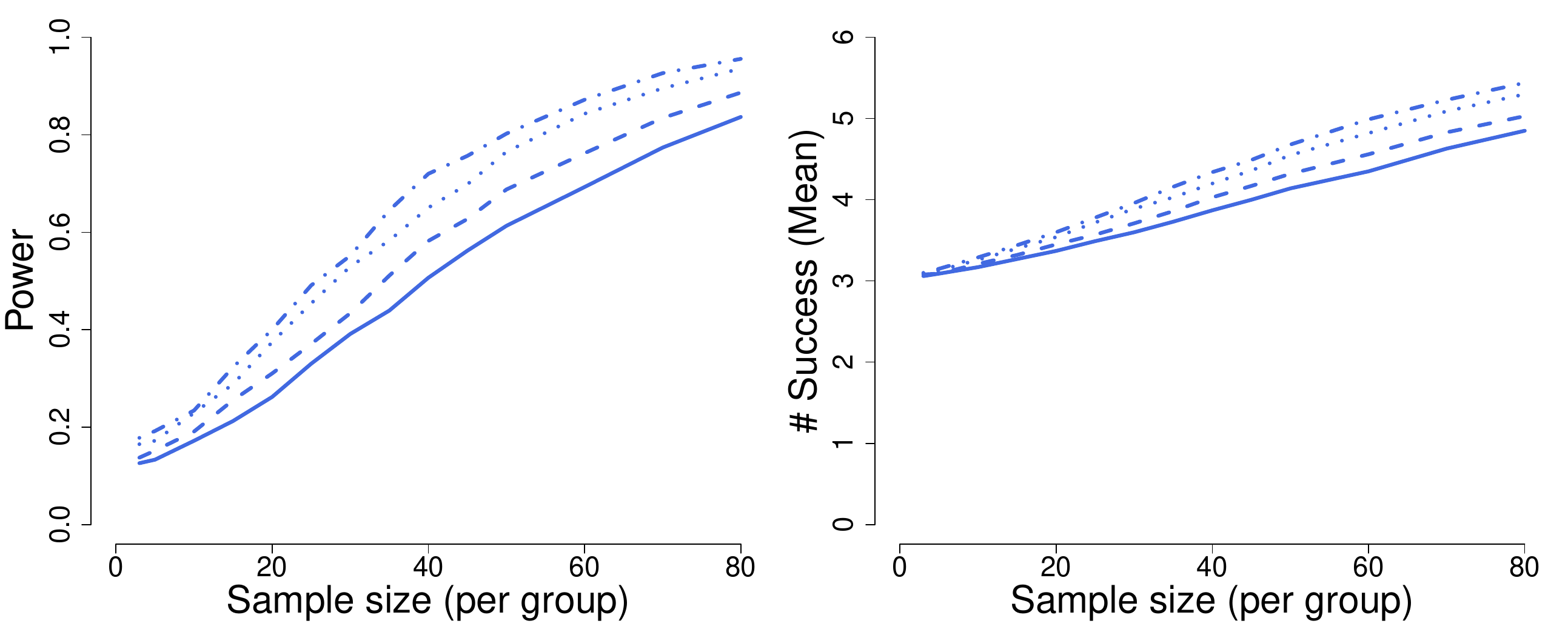}}\\
{\bf Model 2.}\\
{\includegraphics[width=15cm]{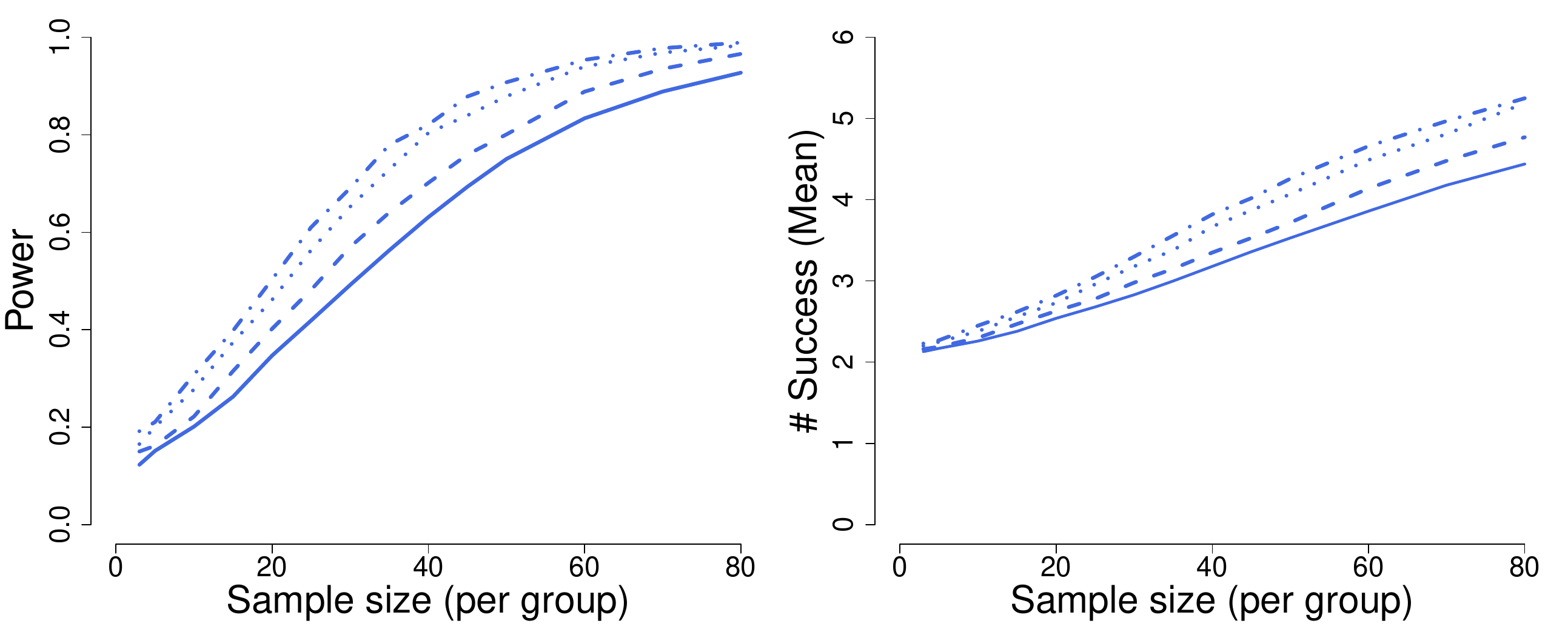}}\\
{\bf Model 3.}\\
{\includegraphics[width=15cm]{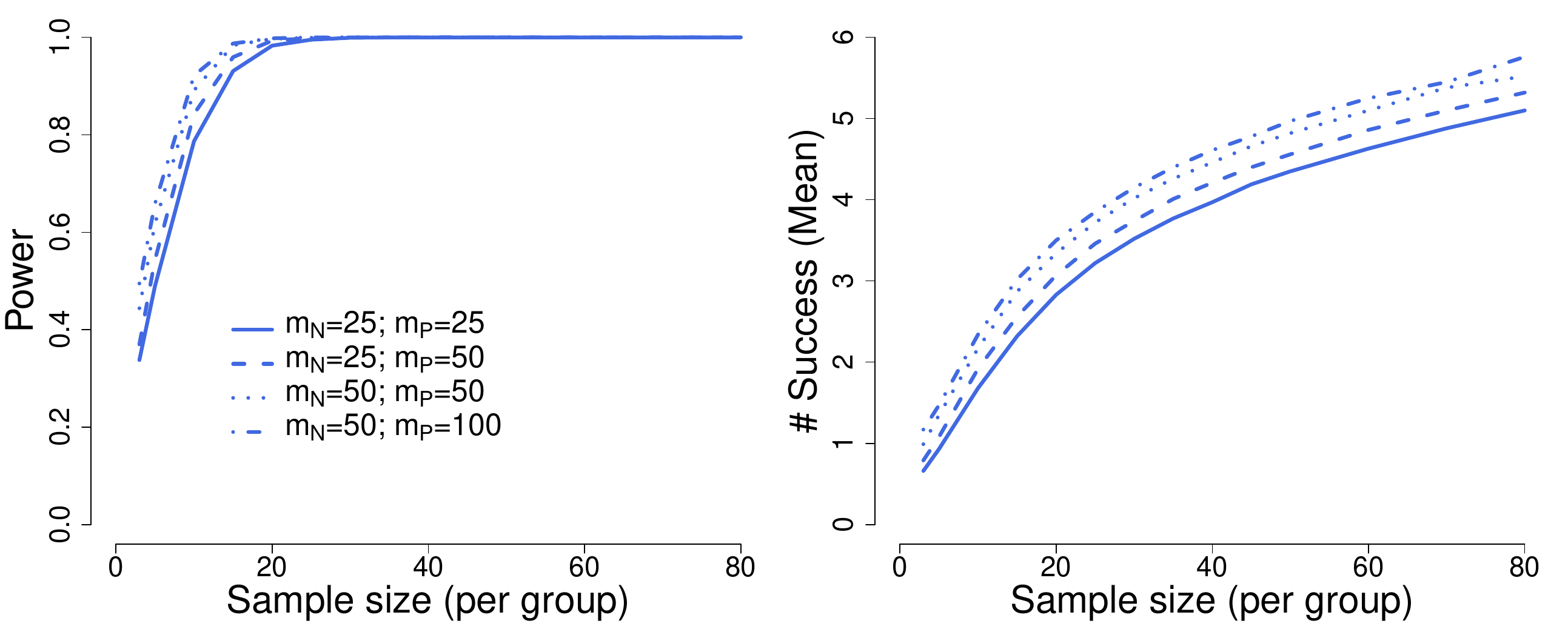}}
\end{tabular}
\end{center}
\caption{{\bf Success (${\boldsymbol {\sigma_\epsilon=0.3}}$)}. Rejection proportions and average number of correctly classified pairs from Models 1, 2 and 3, for the case $\sigma_\epsilon=0.3$. Estimations based on 5,000 Monte Carlo iterations.}
\label{Power1}
\end{figure}

Figure \ref{Power1} shows the evolution of the rejection percentages (left) and the average number of success in the post hoc pairs comparisons (right) for the considered alternatives when $\sigma_\epsilon=0.3$. The three different studied alternatives were,
\begin{align*}
{\text {\bf Model 1.  }}& {\cal A}_1={\cal A}_2={\cal A}_3=0.65;\, {\cal A}_4=0.7.\qquad\qquad\qquad\qquad\qquad\qquad\qquad\\ 
{\text {\bf Model 2.  }}& {\cal A}_1={\cal A}_2=0.65;\, {\cal A}_3={\cal A}_4=0.7.\\ 
{\text {\bf Model 3.  }}& {\cal A}_1=0.65;\, {\cal A}_2=0.7;\, {\cal A}_3=0.75;\, {\cal A}_4=0.8.
\end{align*}

Results are strongly affected by the number of available pre- and post- treatment measures per subject. In the Model 1, the average number of subjects for reaching a power of 0.8 varies between 75 and 50, for the cases $(m_N,m_P)=(25,25)$, and $(m_N,m_P)=(50, 100)$, respectively. In this model, rejecting zero pairs represents an average success of 3. For $n_R=80$, the average number of success increases until 4.85 and 5.44, for  $(m_N,m_P)=(25,25)$, and $(m_N,m_P)=(50, 100)$, respectively. Model 2 shows better results in both, overall and pairs comparison. In this model, no rejecting any pair implies an average success of 2, and in the most powerful scenario, it reached an average of 5.25, $n_R=80$ and $(m_N,m_P)=(50, 100)$. Finally, in the worst case of the Model 3, $(m_N,m_P)=(25,25)$, the power of the overall test is above 0.8 for $n_R=10$. The post hoc test (all pairs are different) also shows a good improvement from the smallest sample size. Final average values ($n_R=80$) ranged between 5.10 to 5.76.

\begin{figure}
\begin{center}
\begin{tabular}{cc}
{\bf Model 1.}\\
{\includegraphics[width=15cm]{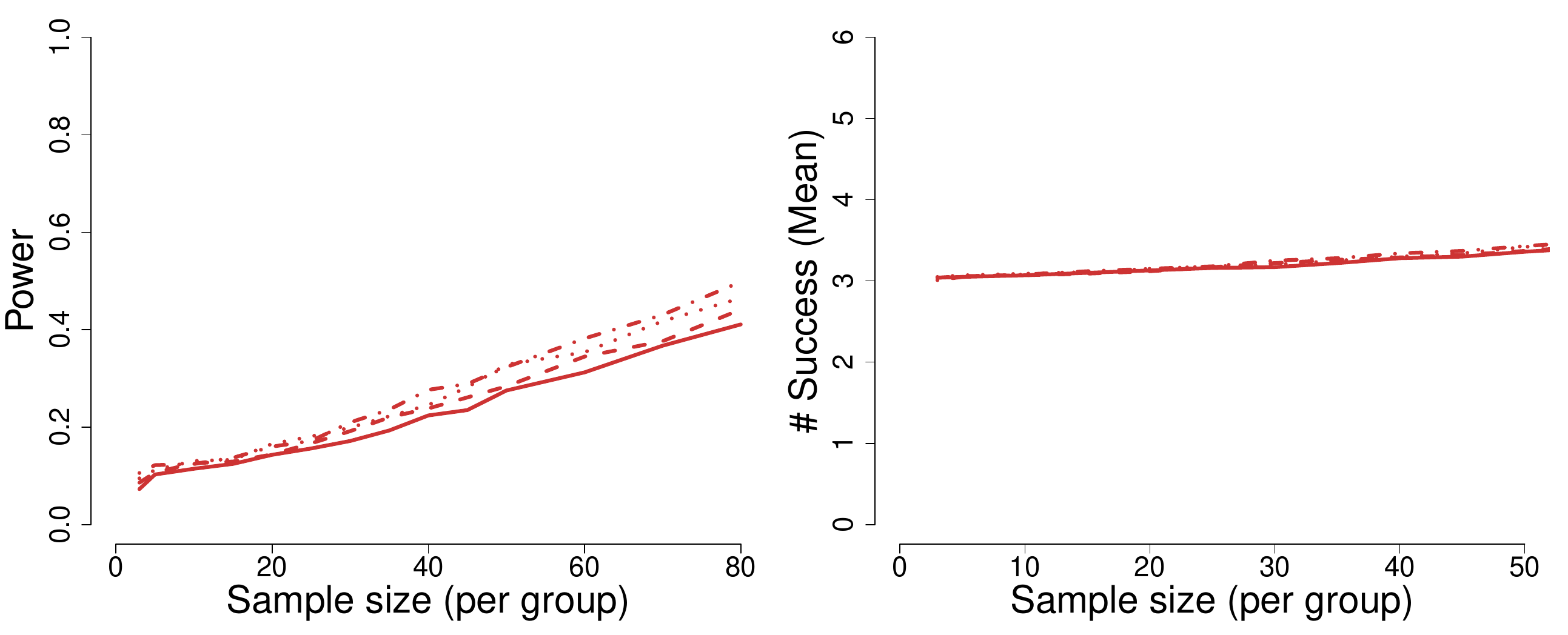}}\\
{\bf Model 2.}\\
{\includegraphics[width=15cm]{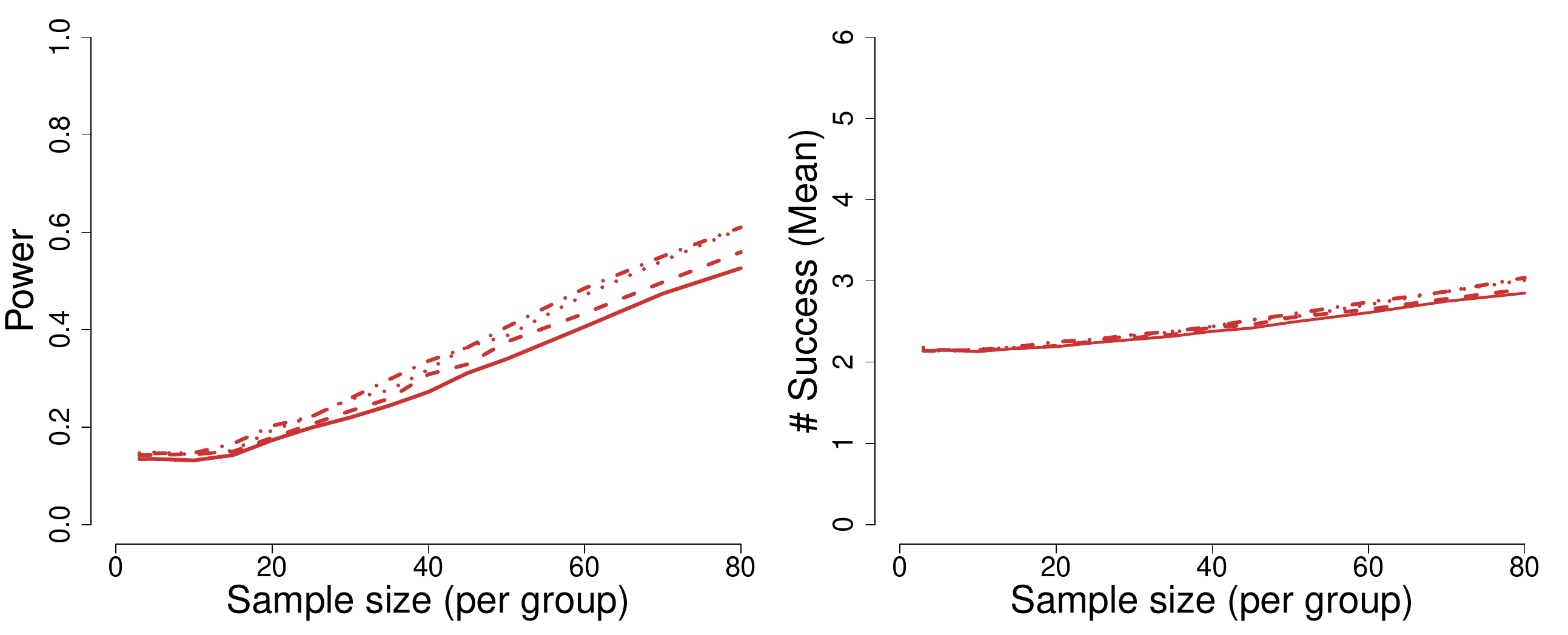}}\\
{\bf Model 3.}\\
{\includegraphics[width=15cm]{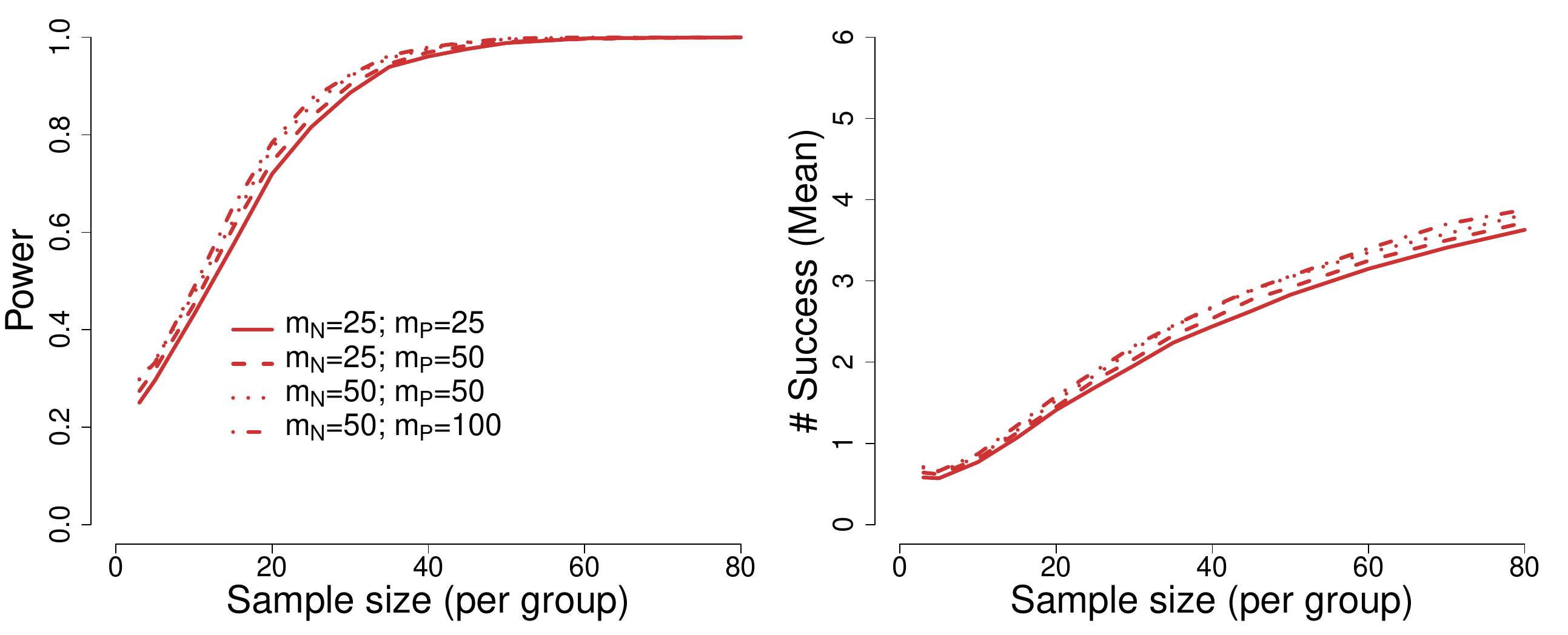}}
\end{tabular}
\end{center}
\caption{{\bf Success (${\boldsymbol {\sigma_\epsilon=0.6}}$)}. Rejection proportions and average number of correctly classified pairs from Models 1, 2 and 3, for the case $\sigma_\epsilon=0.6$. Estimations based on 5,000 Monte Carlo iterations.}
\label{Power2}
\end{figure}

Figure \ref{Power2} is equivalent to the previous one for $\sigma_\epsilon=0.6$. The observed results are also in the same direction, although we can see that the procedure requires larger sample sizes when we have larger heterogeneity between the individual within the same treatment. Besides, the impact of the number of measures per subject in the results is smaller. Not surprisingly, the procedure is little unstable for the very small sample sizes (it can not provide a good estimation of the between-subjects variability). Besides, the power of the test suffers in this case. An $n_R$ of 80 was not enough for getting a power of 80\% in the Model 1, nor in the Model 2. Pairs comparison did not reach neither good averages of success. For $n_R=80$, best average was 3.79 and 3.04, for Model 1 and Model 2, respectively. Model 3 provides better results; sample size required for 80\% of power was around 25. The average number of success in the pairs comparison when $n_R=80$ ranged between 3.63 to 3.88.

\section{Studying the effect of brain stimulation treatments in animal models}

We analyze in this section the problem which motivated this research. Highlighting again that, since the basic research is still in process, presented data have been manipulated. They do not fully depict the observed reality and, therefore, no (real) neuroscience specific conclusions should be extracted from the results here provided.

\subsection{Model construction}

\begin{figure}
\begin{center}
\begin{tabular}{cc}
{\includegraphics[width=14cm]{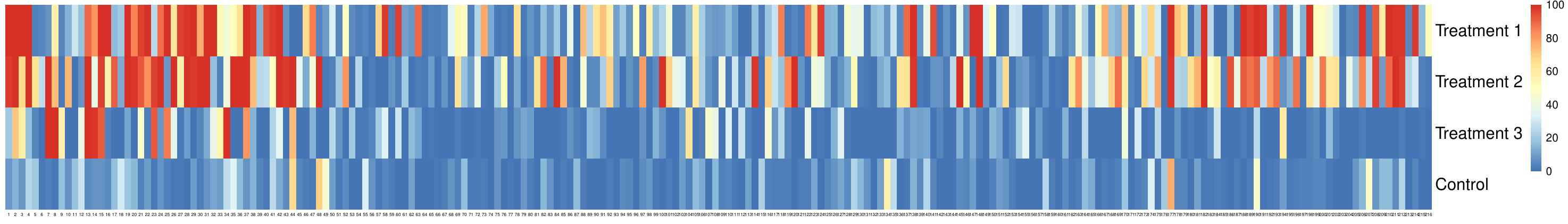}}\\
{\includegraphics[width=14cm]{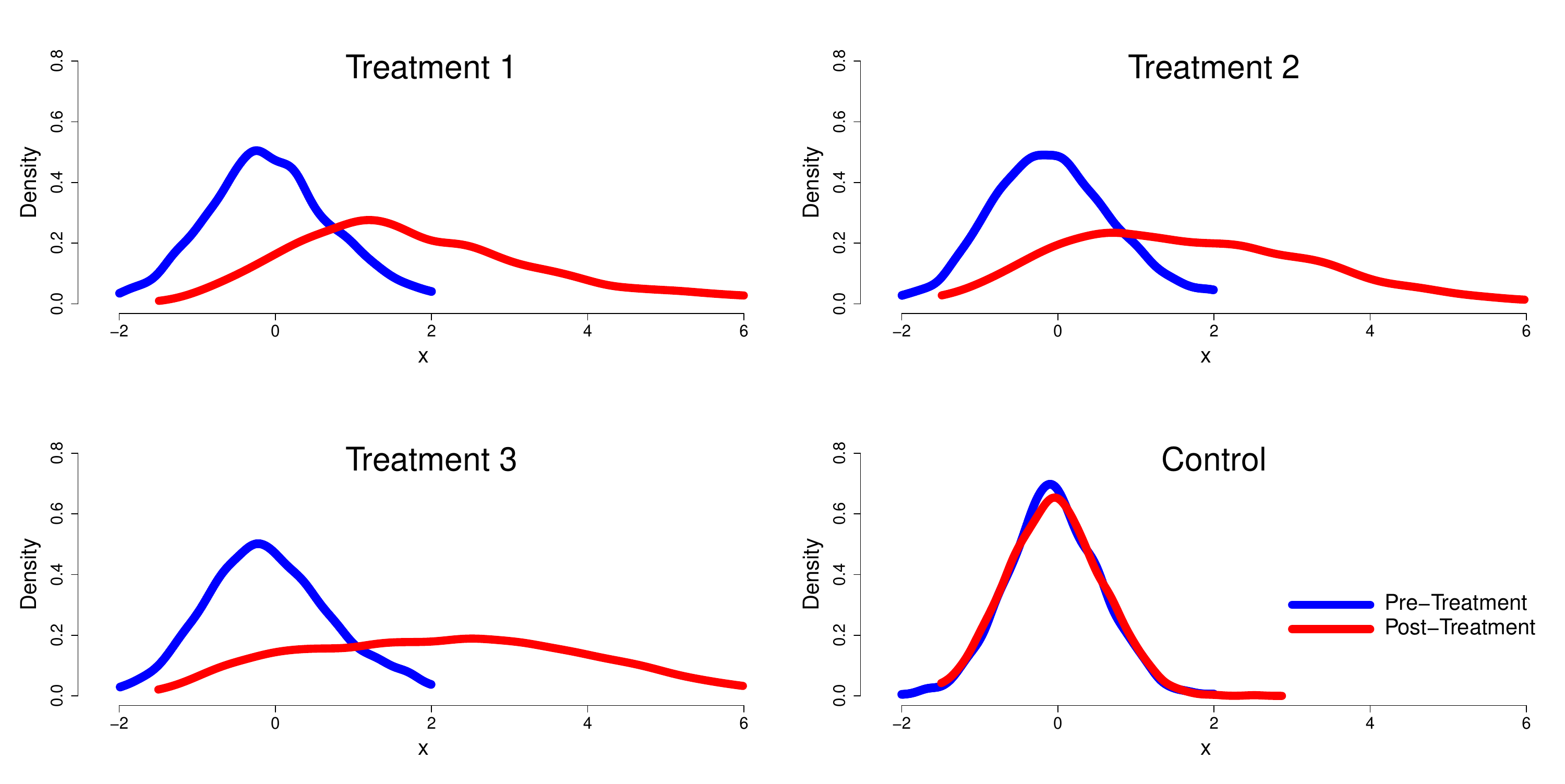}}\\
\end{tabular}
\end{center}
\caption{{\bf Model}. Heatmap with the participation of each feature in the model construction (upper), and kernel density estimations for the final pre- and post- treatment punctuations in the four different treatments (bottom). }
\label{model}
\end{figure}

We received a collection of {\it normalized} (to allow for comparison across days, we normalized power as the percent of total power in a given frequency range and coherence as the coherence in a given frequency range divided by the average coherence across all frequencies) data which include subject and treatment identifiers, and a number of pre- and post- treatment measures (power and coherence) per subject. Since the data contained some noise, first, for each single animal, separately on the pre- and post- treatment measures, and in order to minimize the potential influence of the outliers on the models, we performed a winsorization \cite{dixon60}. The robust Hampel filter criteria \cite{hampel74} (median$\pm 3\cdot$MAD, where MAD is the median absolute deviation from the median) was used for the trimming. Second, for each animal, each measure was standardized based on the distribution of the pre-treatment measures. Then, we randomly selected the 25\% of the sample and, for each treatment, constructed a penalized logistic regression model using the least absolute shrinkage and selection operator (LASSO) for the regularization \cite{tibshirani96}. The \url{R} package glmnet \cite{friedman10} was used for the practical implementation. LASSO approach also allows knowing which of the collected features are more useful for separating the pre- and post- measures in each particular model. Figure \ref{model} (upper) shows the number of times the absolute value of the coefficient associated with each feature is greater than 0.05 in 100 bootstrap iterations of the model construction in the four different treatments. We can see that, while Treatment 1 and Treatment 2 seem to modify similar features, Treatment 3 is mostly impacting in a small number of them, and, not surprisingly, the Control group does not show relevant changes in any of the features. Figure \ref{model} (lower) shows the overall density estimations for the final obtained punctuations on the 75\% of the sample not used in the four models construction in the pre- and post- measures. The differences on both location and variability between the final punctuations in the pre- and post-treatment measures in the Treatments 1-3 are evident. In Treatment 4 (control group) the distributions look similar.

\subsection{Performance comparison}

A total of 100 animals were finally analyzed in the testing cohort, with averages numbers of 149, and 290 pre- and post- treatment measures, respectively. Based on the final model punctuations, the AUCs per rat  ranged between 0.39 and 0.99. Besides, we observed a great impact of the animal (large between-subjects variability).  Averages were similar for treatments 1, 2 and 3 (around 0.85), while the control group behaved as expected, and provided an average AUC of 0.50. Table \ref{description} includes a full description of observed AUCs by treatment, and the sample sizes involved in their computes.

\begin{footnotesize}
\begin{table}
\caption{{\footnotesize {\bf Description.} Full description of observed AUCs, and of the sample sizes involved in their computes by treatment. SD= standard deviation; $n=$ number of animals; $m_P=$ average number of post-treatment measures; $m_N=$ average number of pre-treatment measures. \label{description}}}
\begin{center}
\begin{footnotesize}
\begin{tabular}{lcccccccccccccccccccccccc}
\rowcolor{black!15} 
&& \multicolumn{3}{c}{\bf Areas Under the Curve } & & & &\\
\rowcolor{black!15} 
                          && {\bf Mean$\pm$SD} & {\bf Minimum} & {\bf Maximum} && {$\boldsymbol n$} & {$\boldsymbol {m_P}$} & {$\boldsymbol {m_N}$}\\
{\bf Control}       && 0.50$\pm$0.04 & 0.39 & 0.56 && 21 & 150 & 149\\
\rowcolor{black!5} 
{\bf Treatment 1} && 0.86$\pm$0.08 &  0.68 & 0.99 &&  28 & 201 & 149\\
{\bf Treatment 2} && 0.83$\pm$0.11 & 0.62 & 0.96 && 25 & 209 & 147\\
\rowcolor{black!5}
{\bf Treatment 3} && 0.85$\pm$0.09 & 0.65 & 0.98 && 26 & 601 & 150\\
\end{tabular}
\end{footnotesize}
\end{center}
\end{table}
\end{footnotesize}

\begin{figure}
\begin{center}
{\includegraphics[width=11cm]{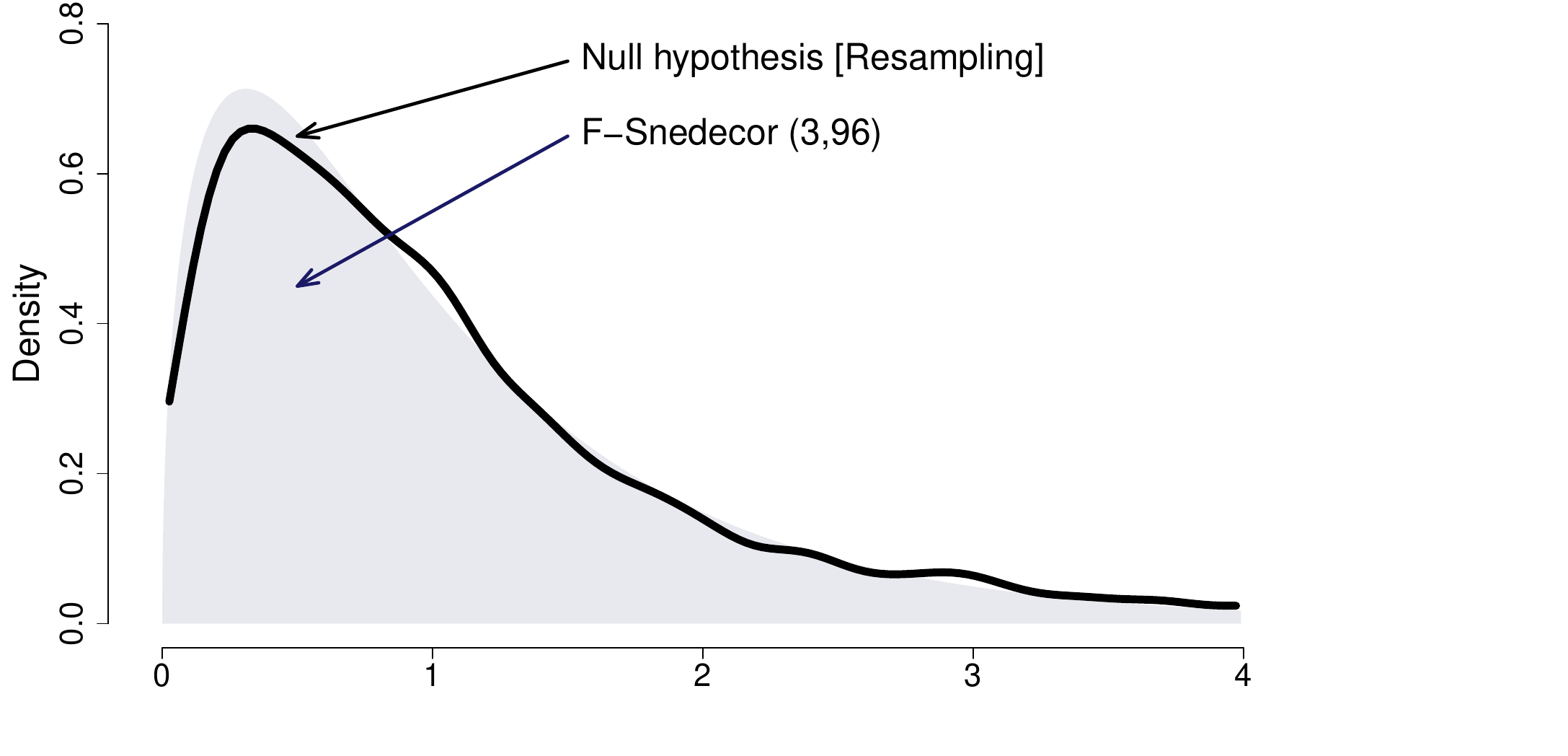}}
\end{center}
\caption{{\bf Distributions.}  Kernel density estimation based on 5000 iterations of a resampling procedure for the brain stimulation data under the null hypothesis, and the density of the expected F-Snedecor.}
\label{distributions}
\end{figure}

We checked how our asymptotic results fit with the problem at hand. We randomly permutated the treatment labels and, within each rat, made a resampling (with replacement) for the pre- and post- treatment measures. Then, we computed the value of the resulting statistics and repeat that 5000 times. Figure \ref{distributions} shows the kernel density estimation for those values, and the density of the expected F-Snedecor distribution (3 and 96 degree of freedom, respectively). The result confirms the good behavior of the proposed approximation for this particular case, despite the observed amount of between-subjects variability.

\begin{figure}[H]
\begin{center}
{\includegraphics[width=16cm]{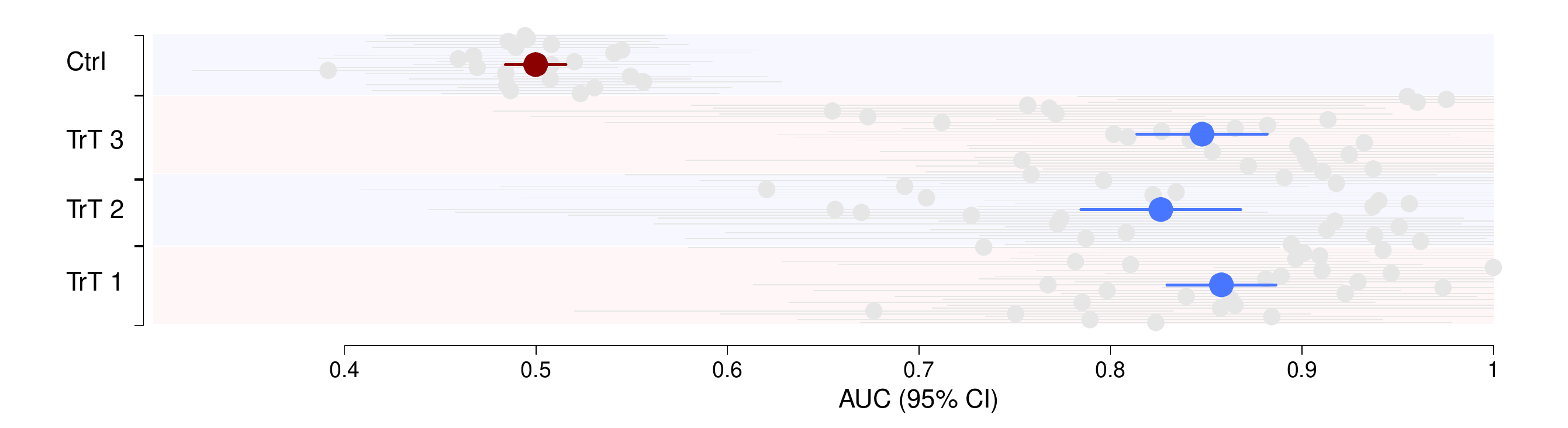}}\\
\phantom b\par
{\includegraphics[width=15cm]{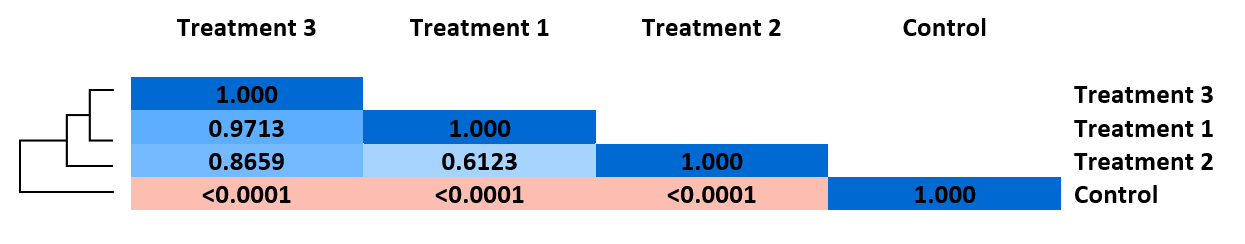}}
\end{center}
\caption{{\bf Results.} Forest plot for the individual and per treatment AUCs with 95\% confidence intervals (up), and a representation of the pair comparisons with a dendrogram based on p-values indicating the best association between the pairs (down).}
\label{results}
\end{figure}

The final value of the F-statistics in our data was 230.5, corresponding with a p-value below 0.0001. We can conclude that the impact of the treatments studied features is not the same. Applying the proposed HSD-type post hoc test, we observe that pairs comparison between treatments 1, 2, and 3 provide high p-values (all of them greater than 0.6), while the pairs comparison between each of those and the Control group reported p-values below 0.0001. Therefore, the overall conclusion could be that the three considered treatments have a significant and quantitatively similar effect but qualitatively different (see Figure \ref{model}). Figure \ref{results} contains a forest plot with the individual and per treatment AUCs (with 95\% confidence intervals), and a representation of the pairs comparison with a dendrogram indicating the best association between the pairs. The p-value associated with the comparison of the three treatments (excluding the 21 animals in the control group) was 0.5053 (F-Snedecor of  0.68878).

\section{Computational considerations}

The procedures proposed in this manuscript are implemented in the R package \url{aovAUC} freely available at \url{https://github.com/perezsonia/aovAUC}. It is easy to use, and one simple sentence directly returns a table with the main results. Main output looks:\par
\phantom b\par\noindent
\begin{small}

{\color{blue} \begin{verbatim} > aovAUC(Values ~ TrT, ID, ph=TRUE, data = dt) \end{verbatim}}

\begin{verbatim}
Call:
Values ~ TrT

             Sum Square DF Mean Square F-Snedecor   p-value    
Intra-group       96.02 96        1.00     230.49 < 2.2e-16 ***
Inter-groups     691.62  3      230.54                         
Total            787.64                                        
---
Signif. codes:  0 `***' 0.001 `**' 0.01 `*' 0.05 `.'  0.1 ` ' 1

Average random-effects standard error of 0.071 (0.075, 0.105, 0.088, 0.016)

----------------------------------------------------------------------

Post hoc test (p-values)

            Treatment 1 Treatment 2 Treatment 3 Treatment 4    
Treatment 1                  0.6142      0.9705   < 2.2e-16 ***
Treatment 2                              0.8638   < 2.2e-16 ***
Treatment 3                                       < 2.2e-16 ***
---
Signif. codes:  0 `***' 0.001 `**' 0.01 `*' 0.05 `.'  0.1 ` ' 1
\end{verbatim}
\end{small}\par
\phantom b\par

More information is obtained through the command \url{summary}. Full information about the package is provided as supplementary material.

\section{Conclusions}

The area under the receiver-operating characteristic curve, AUC, has become a popular index for measuring the size for the association between continuous variables and binary outcomes \cite{camblor20}. In the original study here considered, and which motivated this research, it is used as a metric to represent the difference in brain states induced by an intervention in an animal model. Although some work has been done regarding the  sample sizes required for effective AUCs comparisons \cite{kim14, shu23}, and other related topics \cite{tcheuko16}, traditional statistical tools used for the comparison of treatment through more conventional indices (for instance, location parameters) such as the analysis of variance have not been fully developed for the AUC. The current problem, which involves more than one measure per animal and, therefore, a random-effects component, is the perfect excuse for developing a specific random-effects ANOVA for the AUCs comparison.

The proposed procedure uses the asymptotic properties of the non-parametric (empirical) estimator for the AUC to contrast the equality between the involved AUCs through a balance between the variance between- and within- the considered groups. The finite-sample behavior of the resulting test was appropriate for the models considered in the Monte Carlo simulations, although it could become anti-conservative if the balance between the sample size and the random-effect component is not adequate. Logically, if the effect of the animal is large, and we do not have enough data for a correct estimation of the involved parameters, the computed asymptotic variance can become incorrect. The impact of a large effect of the random component is also reflected in the loss of statistical power observed in the Monte Carlo simulations presented (see supplementary material for additional Monte Carlo simulations). Besides, in a standard way, we also presented and studied a post-hoc test that allows knowing which of the involved pairs are really different (if any), keeping the fixed nominal level, and avoiding the multitesting problem. The HSD-type test actually keeps the nominal level under the null (results shown a little conservative behavior, not unusual in these type of tests) while the number of actually different pairs  detected increased with the sample size.

Regarding the real data analysis presented (Recall: since the basic research is still in process, presented data have been manipulated. They do not fully depict the observed reality and, therefore, no real neuroscience specific conclusions should be extracted from the results here provided), we want to highlight mainly two points. One unrelated with the current research. 1) LASSO regression results could help us not only to identify the separation between pre- and post- measures, but also to know which aspects of the brain activities are more affected by each particular treatment. 2) Quantitatively, the three considered treatments provided good and similar separation between the pre- and post- measures, although they are qualitatively different.

In short, the proposed methodology behaves adequately and we hope that the provided R package will help its use in real practice. Although aspects related with the sample size computation involved a number of parameters, and the meaning of some of the components are not the same, standard procedures based on the portion of explained variance could be used in this context. The analysis of the particularities of this problem, and the development of a MatLab (software frequently used for doing the quantifications of brain activity) are in the {\it to do list} of future related work.

\section*{Funding}
This work was supported from the Grants GRUPIN AYUD/2021/50897 from the Asturies
Government and PID2020-118101GB-I00 from Ministerio de Ciencia e Innovaci\'on (Spanish Government).

\section*{Online supplements}
As online supplementary materials, we provide: 1) A file with the \url{R} code used for implementing the example included in Section 6. 2) The \url{R} package aovAUC, which implements the procedures here proposed, including the documentation of the package, and 3) A document containing additional Monte Carlo simulations.

\section*{Conflict of Interest}
The authors do not have conflict of interest to report.

\section*{Data Availability Statement}
Since the research which motivated this procedure is still in process, we cannot publish the full used data. In the provided package, we include the subset of data used in the final part of the Section 6.

\section*{Appendix: Between-subjects variance estimation}

For estimating the between-subjects variance, we argue as in \citet{kacke04}, and use the equality (for each $i\in \{1,\cdots k\}$)
\begin{align*}
\mathbb E\left[\sum_{j=1}^{n_i} (\hat {\cal A}_{i,j} - {\cal A}_i)^2\right]&= \sum_{j=1}^{n_i} \mathbb E[(\hat {\cal A}_{i,j} - {\cal A}_{i,j})^2 + ({\cal A}_{i,j} - {\cal A}_i)^2 + 2\cdot (\hat {\cal A}_{i,j} - {\cal A}_{i,j})\cdot  ({\cal A}_{i,j} - {\cal A}_i)]\\
&=  \sum_{j=1}^{n_i}\sigma_{i,j}^2 + n_i\cdot\tau^2_{a_{i,\bullet}}.
\end{align*}
Directly applying the general method-of-moments, we obtain the estimator
\begin{align*}
\hat \tau^2_{a_{i,\bullet}}= \frac{1}{n_i-1}\sum_{j=1}^{n_i} (\hat {\cal A}_{i,j} - \hat{\cal A}_i)^2 - \frac{1}{n_i} \sum_{j=1}^{n_i}\hat\sigma_{i,j}^2.
\end{align*}

\bibliographystyle{unsrtnat}
\bibliography{Bibliography}
\end{document}